\documentclass{aa}  
\usepackage{graphicx}
\usepackage{txfonts}
\newcommand{\HI}{H\:\!\textsc{i}}

\begin{document}

   \title{Galaxy metallicities depend primarily on stellar mass and molecular gas mass}


   \author{M. S. Bothwell    \inst{1,2}
          \and
R. Maiolino  \inst{1,2} 
  \and
C. Cicone  \inst{3} 
  \and
Y. Peng  \inst{4,1,2}
  \and
J. Wagg  \inst{5}
          }

   \institute{Cavendish Laboratory, University of Cambridge, 19 J.J. Thomson Avenue, Cambridge, CB3 0HE, UK \\
              \email{matthew.bothwell@gmail.com}
         \and
             Kavli Institute for Cosmology, University of Cambridge, Madingley Road, Cambridge CB3 0HA, UK \\
             \and	
             ETH Zurich, Institute for Astronomy, Wolfgang-Pauli-Strasse 27, CH-8093 Zurich, Switzerland\\
             \and
             Kavli Institute for Astronomy and Astrophysics, Peking University, Beijing 100871, China\\
	      \and
	      SKA Organisation, Lower Withington, Macclesfield, Cheshire,  UK, SK11 9DL\\
             }

   \date{\today}

 
  \abstract
   {}
   {In this work we present an analysis of the behaviour of galaxies in a four-dimensional parameter space defined by stellar mass, metallicity, star formation rate, and molecular gas mass. We analyse a combined sample of 227 galaxies, which draws from a number of surveys across the redshift range $0 < z < 2$ ($>90\%$ of the sample at $z\sim0$), and covers $>3$ decades in stellar mass.}
   {Using Principle Component Analysis, we demonstrate that galaxies in our sample lie on a 2-dimensional plane within this 4D parameter space, indicative of galaxies that exist in an equilibrium between gas inflow and outflow. Furthermore, we find that the metallicity of galaxies depends only on stellar mass and molecular gas mass. In other words, gas-phase metallicity has a negligible dependence on star formation rate, once the correlated effect of molecular gas content is accounted for.}
   {The well-known `fundamental metallicity relation', which describes a close and tight relationship between metallicity and SFR (at fixed stellar mass) is therefore entirely a by-product of the underlying physical relationship with molecular gas mass (via the Schmidt-Kennicutt relation).}
   {}


   \maketitle
%

\section{Introduction}

The abundance of heavy elements within the interstellar medium (ISM) of galaxies remains one of the most important diagnostics of the galaxy evolution process. Heavy elements are produced by supernovae, and the metallicity (the abundance of these heavy elements relative to hydrogen, traced via the oxygen to hydrogen ratio O/H) is affected by both gas inflows and outflows. The metallicity of galaxies therefore effectively functions as a `fossil record' of the physical processes driving their past evolution. Attaining a comprehensive understanding of the factors affecting the metal abundance of galaxies is therefore a critical goal of galaxy evolution studies. 

It has long been known that metallicity correlates with a host of physical parameters -- the most well-known being the correlation with stellar mass (the `mass-metallicity' relation), whereby the stellar mass and gas-phase metallicity of galaxies are closely correlated across a wide range of masses (e.g., \citealt{2004ApJ...613..898T}; \citealt{2006ApJ...647..970L}), and out to high redshift (e.g., \citealt{2005ApJ...635..260S}; \citealt{2008A&A...488..463M}). The existence of the mass-metallicity relation has become a central pillar of our understanding of galaxy evolution, and a key observable for theoretical models to reproduce. In recent years, authors have sought to address secondary correlations in the mass-metallicity relation. \cite{2010MNRAS.408.2115M} and \cite{2010A&A...521L..53L} both found a significant secondary dependence on star formation rate (SFR), such that galaxies at a fixed stellar mass displayed an inverse correlation between metallicity and SFR. Known as the `Fundamental Metallicity Relation' (FMR), in this formalism galaxies can be pictured as lying on a 2D plane in the 3D parameter space defined by stellar mass, metallicity, and star formation rate. The mass-metallicity relation, and the SFR-M$_*$ `main sequence' are therefore best understood as being projections of this underlying 3D distribution along either the SFR axis, or the metallicity axis (respectively). Interestingly, the existence of the FMR naturally explains the observed redshift evolution of the mass-metallicity relation: galaxies at high redshifts have elevated SFRs, relative to $z\sim0$, and these higher SFRs result in concomitantly lower metallicities. Galaxies at high-$z$ are not discrepant from the local mass-metallicity relation; rather, they lie on a different region of the FMR.

The discovery of the interrelation between stellar mass, metallicity, and SFR has prompted extensive followup efforts in order to understand the physics driving these observed correlations. Theoretical models suggest that these relations result from galaxies existing in an equilibrium between gas inflows and gas outflows (\citealt{2010MNRAS.408.2115M}; \citealt{2012arXiv1202.4770D}; \citealt{2012MNRAS.427..906H};  \citealt{2013ApJ...772..119L}; \citealt{2013arXiv1302.3631D} \citealt{2014arXiv1404.0043O}), while some observational results suggested that the FMR may in fact be more strongly expressed via gas content than SFR (\citealt{2013MNRAS.433.1425B}). \cite{2014ApJ...791..130Z} derive a model whereby the FMR is best understood as a byproduct of an underlying relationship between metallicity and gas fraction: this model is supported by the results presented by \cite{2016MNRAS.455.1156B}, who demonstrate observationally that the SFR-FMR is likely to be a by-product of a more fundamental relation between stellar mass, metallicity, and molecular gas mass.

In this work, we present a new 4-dimensional principle component analysis of a large sample of galaxies lying between $0 < z < 2$. Previous observational analyses undertaken with the aim of studying the FMR have been hampered by the degeneracy between SFR and M(H$_2$), which are linked via the star formation law. In \cite{2016MNRAS.455.1156B} we compared the `SFR-FMR' (the 3D relation between stellar mass, metallicity, and SFR) with the `H$_2$-FMR' (the 3D relation between stellar mass, metallicity, and M(H$_2$)). In this work we directly compare the effects of both SFR and M(H$_2$) on the mass-metallicity relation, by carrying out a 4-dimensional PCA. The use of a 4-dimensional PCA allows us to break the degeneracy between SFR and M(H$_2$), and uncover the primary driving mechanism behind the fundamental metallicity relation. We demonstrate that the observed SFR-FMR is indeed entirely a by-product of the true underlying relation between metallicity and molecular gas, with the SFR-FMR emerging as a result of the combination of (a) the metallicity-gas mass correlation, and (b) the star formation law. Indeed, once the correlated effect of molecular gas mass has been removed, the metallicity shows essentially zero remaining dependence on star formation rate. We discuss our sample selection in \S2, and give our results in \S3. We discuss these results in \S4, and conclude in \S5. Throughout, we adopt a cosmology following \cite{2015arXiv150201589P}, and a \cite{2003PASP..115..763C} IMF.

\section{Sample selection}
\label{sec:sample}


\begin{figure*}
\centering
\includegraphics[width=10.5cm]{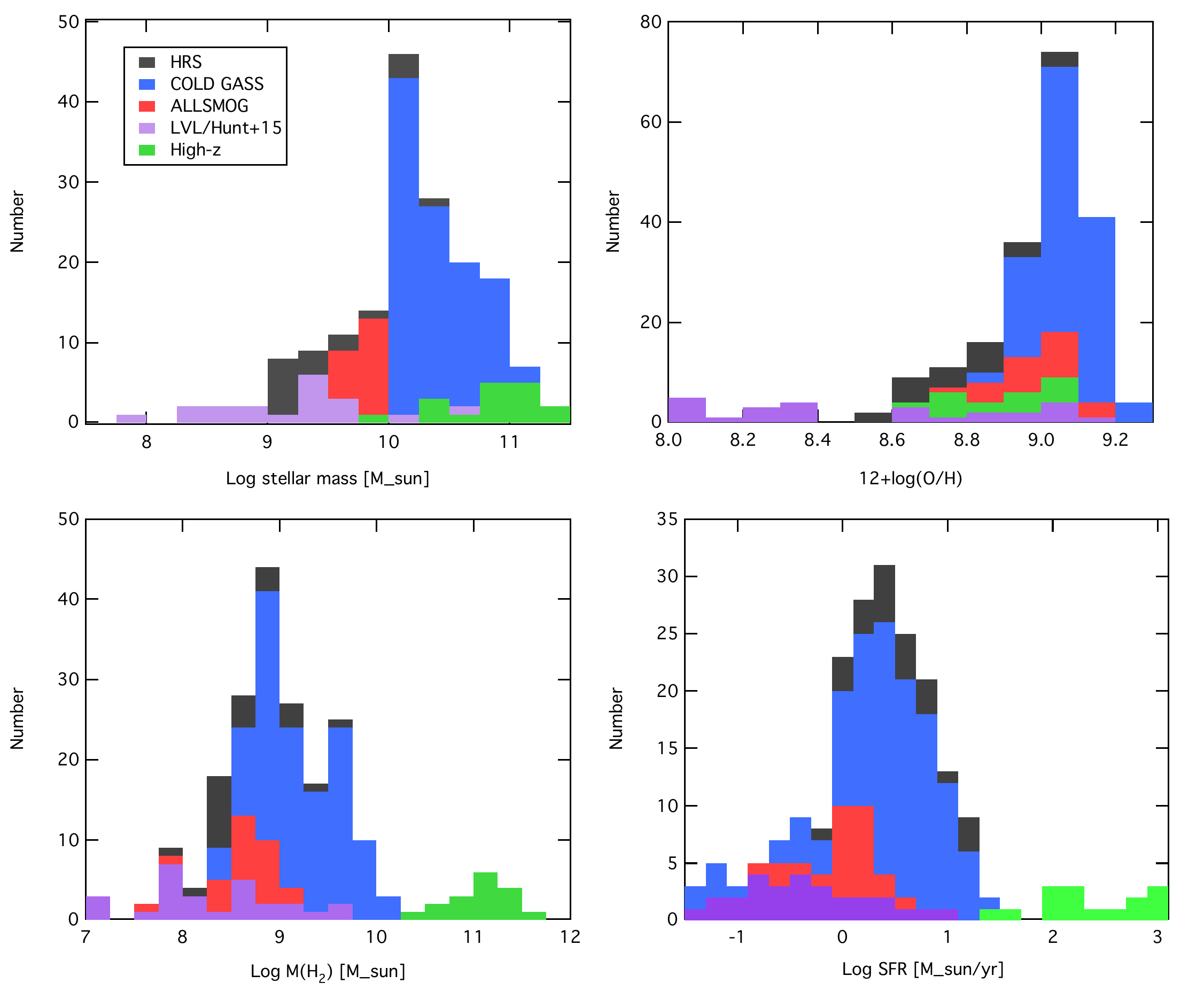}	
\caption{Histograms of the stellar mass ({\it upper left panel}), metallicities ({\it upper right panel}), star formation rates ({\it lower left panel}), molecular gas masses ({\it lower right panel}), of galaxies in the samples used in this work. Bars are shown as stacks, differentiating the different samples.}
\label{fig:hist}
\end{figure*}

%
Our sample draws from a number of surveys in both the local and high-$z$ Universe. Locally, we draw galaxies from the ALLSMOG \citep{2014MNRAS.445.2599B}, COLD GASS \citep{2011MNRAS.415...32S}; Hershel Reference Survey (HRS; \citealt{2010PASP..122..261B}) and Local Volume Legacy (LVL;  \citealt{2009ApJ...706..599L}; \citealt{2010ApJ...715..506M}) surveys. We also include optically-selected `main-sequence' galaxies at high redshift selected from the PHIBBS survey \citep{2013ApJ...768...74T}, and luminous sub-millimetre galaxies (\citealt{2005ApJ...622..772C};  \citealt{2013MNRAS.429.3047B}). Overall, our sample selection was driven by the need to have a range of available physical parameters: namely, the stellar mass, star formation rate, gas-phase metallicity, and molecular gas mass. The simultaneous availability of these latter two parameters was generally the limiting factor defining inclusion into our final sample.
 
We also include a sample of 8 low-metallicity dwarf galaxies presented by \cite{2015arXiv150904870H}. These galaxies have molecular gas masses measured via IRAM 30m observations of their $^{12}$CO($1-0$) emission line, SFRs measured using a combination of H$\alpha$ and $24\mu$m luminosities, and have metallicities in the range $7.7 < $ 12+log(O/H) $< 8.4$ (making them amongst the lowest-metallicity galaxies with CO detections). The inclusion of these 8 galaxies serves to increase the range of metallicities probed by our analysis. We re-derive stellar masses for the galaxies presented by \cite{2015arXiv150904870H}, as the original stellar masses are derived using $3.4 \mu$m luminosities which are potentially contaminated by hot dust emission. We derive stellar masses using archival 2MASS H-band luminosities and $(B-V)$ optical colours, following the `mass-to-light' method of \cite{2001ApJ...550..212B}. This re-derivation lowers the \cite{2015arXiv150904870H} stellar masses by a factor of $\sim 3$. 

Throughout this work, we derive molecular gas masses using the metallicity-dependent CO/H$_2$ conversion factor presented by \cite{2010ApJ...716.1191W}.  The \cite{2010ApJ...716.1191W} factor depends only weakly on radiation field intensity and gas density, scaling as $\ln(\chi/n)$. Following Bolatto et al. (2013) we have taken $\chi/n=1 \times 10^{-2}$, while order-of-magnitude changes in $\chi/n$ result in changes to $\alpha_{\rm CO}$ of $\sim 10-20\%$. Our results are robust to a range of metallicity-dependent CO/H$_2$ conversion factors (i.e., \citealt{2012ApJ...747..124F};  \citealt{2011MNRAS.412..337G}; \citealt{2012MNRAS.421.3127N}, in addition to our chosen prescription, \citealt{2010ApJ...716.1191W}), though the use of conversion factor prescriptions with very steep ($n>2$) power law dependences on metallicity (i.e., \citealt{1997A&A...328..471I}; \citealt{2012AJ....143..138S}) may alter our results.

 We derive metallicities for our low-$z$ samples by taking the mean of the metallicity derived via the R23 and [NII]/H$\alpha$ tracers, using the calibration of \cite{2008A&A...488..463M}. At high redshift, with fewer available optical lines, we derive metallicities using the [NII]/H$\alpha$ tracer alone. Our sample selection is identical to that presented in \cite{2016MNRAS.455.1156B}, and we refer readers to that work for further detail. Our low-$z$ samples, and high-$z$ main sequence galaxies, end up with typical values of $\alpha_{\rm CO}$ in the range $2.5 - 4.5$. High-$z$ SMGs are dynamically turbulent systems with values of $\alpha_{\rm CO}$ driven more by the kinematics of their ISM than their metallicity. We have adopted a conventional value of $\alpha_{\rm CO} =0.8$ for all SMGs (as we discuss below, adopting instead a metallicity-dependent $\alpha_{\rm CO}$ for SMGs does not significantly affect our results). 

Our final, combined sample of galaxies consists of 227 members, spanning a redshift range $0 < z < 2$, though the majority of the sample ($>90\%$) are drawn from local surveys ($z\sim0$). Figure \ref{fig:hist} shows the properties of our combined sample, in the form of histograms of stellar mass, metallicity, star formation rate, and molecular gas mass. 

\section{Analysis and results}
\label{sec:results}

In \cite{2016MNRAS.455.1156B}, we present a 3-dimensional principle component analysis of our data, analysing the stellar mass, metallicity and a third parameter of interest (in turn, the molecular gas mass, the {\it total} gas mass, the star formation rate, and the star formation efficiency). We found that the 3D relation between stellar mass, metallicity, and molecular gas mass to be stronger -- and therefore likely to be more fundamental -- than the 3D relation defined by stellar mass, metallicity, and star formation rate (which is equivalent to the `Fundamental Metallicity Relation', as presented by \citealt{2010MNRAS.408.2115M} and \citealt{2010A&A...521L..53L}). As such the driver of the FMR was most likely the molecular gas content.

Here, we improve upon the results presented by \cite{2016MNRAS.455.1156B} in two ways. Firstly, by including additional low-mass galaxies (taken from observations by \cite{2015arXiv150904870H}, and additional ALLSMOG galaxies). Secondly, and most importantly, for the first time we directly perform a single, simultaneous 4-dimensional principle component analysis on the four parameters stellar mass, metallicity, star formation rate, and molecular gas mass.

Principle Component Analysis (PCA) is a parameter transformation technique, whereby a set of physical parameters are converted into a set of orthogonal (and linearly uncorrelated) vectors, or `Principle Components'. The transformation is defined such that the maximum amount of variance is contained within the first principle component, and then each subsequent component contains as much remaining variance as possible (with the constraint that every component remains orthogonal). In practice, PCA performs a coordinate transformation which (a) reveals the optimum `projection' of a dataset, and (b) reveals which parameters are responsible for the variance in the sample. PCA is particularly useful for revealing any possible reduction in dimensionality -- for example, by revealing that some dataset lies on a 2D `plane' in 3D parameter space. PCA was used by both \cite{2010A&A...521L..53L} and \cite{2012MNRAS.427..906H} to examine the mass-metallicity relation's secondary dependence on SFR.

We first normalise each parameter to the mean value for our combined sample:

$$ \log {\rm (M_*)^{PCA}}=      \log {\rm (M_*)} - 10.12$$
$${\rm 12+log(O/H)^{PCA}}=   {\rm 12+log(O/H)}  -  8.95$$
$$\log {\rm (SFR)^{PCA}}=  \log ({\rm SFR}) -    0.40$$
$$\log {\rm (M_{H2})^{PCA}}= \log {\rm (M_{H2})} -     9.04$$

\begin{figure*}
\centering
\includegraphics[width=12cm]{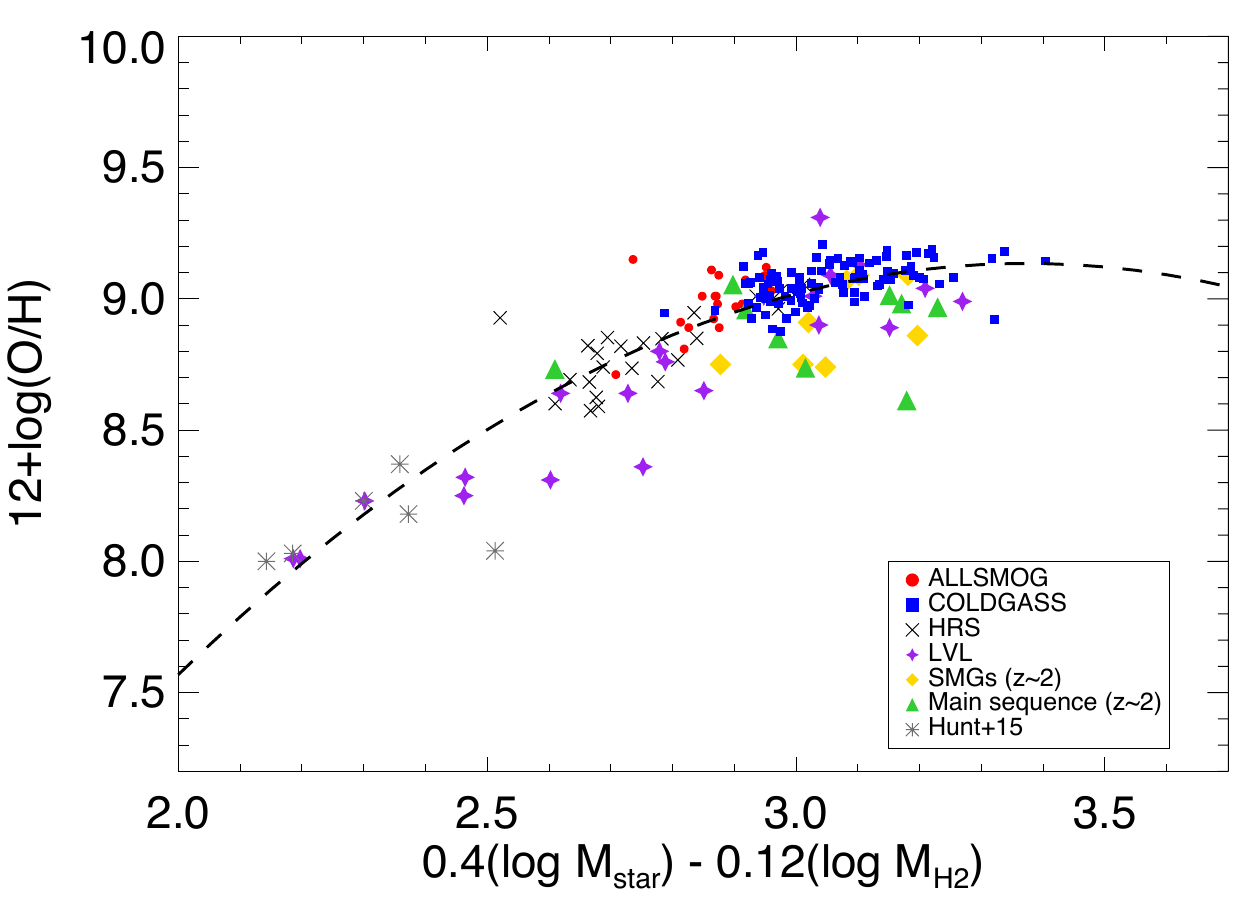}	
\caption{Our fourth Principle Component, plotted as metallicity (12+log(O/H)) vs the optimum linear combination of stellar mass, and molecular gas mass. We have not included the contribution from  SFR, as it is essentially zero: the optimum projection of the data requires only the stellar mass and molecular gas mass. The quadratic fit, defined in Equation 1, is overplotted as a dashed line. Galaxies with only upper limits on their H$_2$ mass are not shown (to avoid crowding the plot), but are consistent with the derived relation.}
\label{fig:FMR}
\end{figure*}

We have accounted for uncertainty by performing a Monte-Carlo bootstrap, performing $10^5$ PCA iterations. During each iteration, each galaxy has the its physical parameters randomly perturbed by an amount defined by the respective error on each parameter. After performing $10^5$ iterations, we take our `final' Eigenvector values (and uncertainties) to be the mean (and standard deviation) of the resulting Eigenvector distribution. In order to ensure we are not unduly influenced by outliers, we also perform sample bootstrapping: during each iteration, we randomly sample with replacement our complete sample of galaxies, generating for each iteration a new sample, with a size equal to our original dataset.

It is important to note that the application of PCA to our data has a potential weakness: the fact that PCA can only describe datasets in terms of linear relationships between parameters. More complex, non-linear distributions of data cannot be described in terms of a simple set of orthogonal eigenvectors. While linear correlations (such as the SFR-M$_*$ `galaxy main sequence') are easily described by PCA, applying PCA to non-linear relations (such as the mass-metallicity relation) will, by definition, be somewhat inaccurate. The practical effect of this will be to increase the apparent scatter around the component vectors. Given the relatively low number of galaxies in our sample, it is likely that the uncertainty added by describing our dataset in terms of purely linear relations is not larger than the scatter inherent in the distribution of data (which our Monte-Carlo technique is designed to reveal). We therefore simply caution the reader that the results revealed by PCA may have some uncertainty added added due to non-linearity in some underlying correlations. 

In addition, we note that while we include galaxies drawn from samples of various sizes, we do not apply any re-weighting based on the parent sample; weighting by sample size would unfairly privilege members of smaller samples (there is nothing inherent about being drawn from a smaller sample that would require a galaxy to be given higher weighting), and weighting by the neighbouring density of points would give disproportionate weighting to outliers. We therefore weight each individual galaxy equally. 

We find that our PCA results demonstrate the existence of a `fundamental plane' in the 4-dimensional parameter space defined by stellar mass, metallicity, star formation rate, and molecular gas mass. The first two 4-vectors (defining a 2D plane) are together responsible for 93\% of all the variance in the data.

The four principle components defining our combined dataset are:

$$ {\rm PC}_1 = 0.37{\rm (log \; M_*)} + 0.67 {\rm (log \; SFR)}- 0.62 {\rm (log \; M_{H2})}- 0.09 (Z)$$
$$ {\rm PC}_2 = -0.79{\rm (log \; M_*)} + 0.52 {\rm (log \; SFR)} - 0.03 {\rm (log \; M_{H2})} - 0.29 (Z)$$
$$ {\rm PC}_3 = -0.27 {\rm (log \; M_*)}- 0.49  {\rm (log \; SFR)} + 0.72 {\rm (log \; M_{H2})}- 0.18 (Z)$$
$$ {\rm PC}_4 = 0.34{\rm (log \; M_*)} - 0.009 {\rm (log \; SFR)} - 0.10 {\rm (log \; M_{H2})} - 0.86 (Z)$$


77\% of the sample variance is contained within $ {\rm PC}_1$, 93\% is contained within ($ {\rm PC}_1 +  {\rm PC}_2$), and 98\% of the sample variance is contained within ($ {\rm PC}_1 +  {\rm PC}_2 + {\rm PC}_3$). The fourth principle component, $ {\rm PC}_4$, is therefore essentially zero (to within 2\%). The fact that the vast majority (93\%) of the sample variance is contained within two vectors shows that the distribution of all galaxies in our sample do indeed lie on a plane in this 4D space, with just 7\% of the sample variance taking the form of scatter around this plane.

As ${\rm PC}_4$ is (a) essentially zero, and (b) dominated by metallicity, we can therefore `solve' for metallicity, by setting the fourth principle component equal to zero (which, as above, is valid at the $\sim 2\%$ level), allowing us to write an expression for metallicity in terms of the other physical parameters.

Setting ${\rm PC}_4$ to zero therefore gives us the optimal projection of the combined dataset: 


\begin{equation}
{\rm Metallicity} =  0.4 \;{\rm (log \; M_*)} - 0.12 \;{\rm (log \; M_{H2})}  \\ -0.01\; {\rm (log \; SFR)} $$
\end{equation}


This tells us that the gas-phase metallicity of galaxies in our sample is determined primarily by the stellar mass (i.e., the well known mass-metallicity relation), with a secondary dependence on the molecular gas mass (the effect of which is $\sim30\%$ as strong as the stellar mass dependence), and a vanishingly small dependence on SFR (the effect of which is  $\sim 2\%$ as strong as the stellar mass dependence). This result is only slightly affected by varying our assumptions as discussed above; adopting a metallicity-dependent  CO/H$_2$ conversion factor for the $z\sim2$ SMGs rather than a constant $\alpha_{\rm CO} =0.8$, results in the expression ${\rm Metallicity} =  0.4 \;{\rm (log \; M_*)} - 0.14 \;{\rm (log \; M_{H2})}  \\ -0.03\; {\rm (log \; SFR)} $; i.e., a slightly higher SFR dependence than given in Eq. 1, but leaving the underlying result unchanged. Likewise, we can examine the effect of removing `starburst' galaxies (systems which may not be in a current equilibrium between inflows/outflows/SF); removing starburst galaxies, defined here as $\tau_{\rm dep}$ (=MH$_2$/SFR) $< 2\times 10^8$yr, results in an optimum projection ${\rm Metallicity} =  0.4 \;{\rm (log \; M_*)} - 0.11 \;{\rm (log \; M_{H2})}   -0.02\; {\rm (log \; SFR)} $.

We have plotted the optimum projection of our data (Eq. 1) in Fig. \ref{fig:FMR}. We fit a quadratic function to the data, finding that the metallicity of all galaxies in our sample follows:

\begin{equation}
{\rm 12+log(O/H)} =   (-0.83 \pm 0.11){\rm \xi^2} + (5.60 \pm 0.61) {\rm \xi} -  (0.31 \pm 0.02),
\end{equation}
where $\xi = 0.4 \;{\rm (log \; M_*)} - 0.12 \;{\rm (log \; M_{H2})}$. \\

Readers should note that Eq. 2 gives the best representation of our dataset. Eq. 1 shows the optimum {\it projection} of the data (close to a 2D plane in the 4D parameter space), but of course the data do not have to follow a linear trend on this plane. 
To avoid overcrowding  Fig. 2, we do not plot the positions of galaxies not detected in CO (which therefore have only limits on their molecular gas masses). We do however stress that these limits are consistent with the relation. We also note that the high-$z$ galaxies seem to lie slightly below the relation; while this could be due to uncertainty in the metallicity measurements at high-$z$, it could also hint at evolution in the physics driving the FMR. Such analysis is beyond the scope of this letter, however. 

Readers will note that the projection of the data shown in Fig. 2 is similar in form to the mass-metallicity relation, with a second order correction for gas mass. The new result we report here is that the star formation rate is not needed to find the optimum projection of the data. {\bf Our principle component analysis has led to the surprising result that the gas-phase metallicity has a negligible dependence on star formation rate, once the correlated effect of molecular gas content is accounted for.}  

The well-known `fundamental metallicity relation', which describes a close and tight relationship between metallicity and SFR (at a given stellar mass) is therefore entirely an incidental by-product of the underlying physical relationship with molecular gas mass (which is linked to SFR via the Kennicutt-Schmidt star formation law).  

\section{Total gas mass}
\label{sec:mtot}

Given our finding the observed `fundamental metallicity relation' is driven by an underlying physical connection between stellar mass, metallicity, and molecular gas content, it is important to discover whether the most fundamental gas component is indeed the molecular gas mass, or whether the total gas mass (\HI\ + H$_2$) will be more important. 

We perform the same PCA exercise as above, replacing the molecular gas mass with the total gas mass, M(gas) = $1.36 \times ({\rm M_{HI} + M_{H2}}$), where the factor of 1.36 is a correction for interstellar helium. 21cm \HI\ observations are available for all of the low-redshift galaxies. \HI\ is not detectable at high redshifts, and it is generally assumed that the ISM of high-redshift star-forming galaxies has only a negligible \HI\ component. For our high-$z$ galaxies, we have therefore assumed M(gas) = M(H$_2$).

This analysis reveals a distribution with more scatter than that with molecular gas alone; 69\% of the sample variance is contained within $ {\rm PC}_1$, 88\% is contained within ($ {\rm PC}_1 +  {\rm PC}_2$), and 98\% of the sample variance is contained within ($ {\rm PC}_1 +  {\rm PC}_2 + {\rm PC}_3$). The data do lie on an approximate plane in the 4-dimensional parameter space defined by stellar mass, metallicity, star formation rate, and total gas mass, with 12\% of the sample variance taking the form of scatter around this plane. Performing the same re-arrangement of the fourth principle component as above, we reach an expression for metallicity

$${\rm Metallicity} =  0.33 \;{\rm (log \; M_*)} - 0.02 \;{\rm (log \; M_{gas})}  \\ -0.03\; {\rm (log \; SFR)}. $$

The strong correlation between molecular gas mass and metallicity is not present between metallicity and total  (\HI\ + H$_2$) gas mass. This can be explained as being due to the fact that gas-phase metallicity, being measured via optical nebular emission lines, is observed in star-forming regions of galaxies. These same regions will be rich in molecular gas (as stars form in molecular clouds). Atomic hydrogen, conversely, can exist in a vast halo reaching far beyond the central star-forming regions of the galaxy. A gas accretion event that increases M(\HI) will not necessarily elevate the star formation rate, or dilute the observed metallicity. It is therefore unsurprising that including the atomic gas component serves to weaken the correlations observed between metallicity and molecular gas mass.

\section{Conclusions}
\label{sec:conc}

We have presented a four dimensional principle component analysis on a sample of 227 galaxies, ranging from low-metallicity dwarfs to massive starbursts, lying at redshifts $0 < z < 2$. Our sample was selected in order to have a full complement of stellar mass, metallicity, molecular gas mass, and star formation rate data. Our 4D PCA has revealed two main conclusions:

\begin{itemize}

\item{Our sample of galaxies lie on a 2D plane in the four-dimensional parameter space defined by stellar mass, metallicity, molecular gas mass, and star formation rate, with 93\% of all the sample variance being contained within the first two principle components.  } 

\item{Setting the fourth principle component to zero, we find an expression for metallicity in terms of stellar mass, star formation rate, and molecular gas mass. We find that the star formation rate has a negligible effect on the metallicity of galaxies in our sample. As such, we conclude that the strong SFR-metallicity relation at a given stellar mass (the `fundamental metallicity relation') is entirely a by-product of the true, physical relation between metallicity and molecular gas content.}

\end{itemize}

\section*{Acknowledgments}

This research has made use of NASA's Astrophysics Data System. ALLSMOG data are available at http://www.mrao.cam.ac.uk/ALLSMOG/. This work is supported by STFC grants ST/M001172/1 and ST/K003119/1.

\end{document}